\RequirePackage{fix-cm}
\documentclass{svjour3}                     
\smartqed  
\usepackage{graphicx}
\begin{document}

\title{The Energy-Momentum Tensor for a Dissipative Fluid in General Relativity
}


\author{Oscar M. Pimentel         \and
        F. D. Lora-Clavijo \and
        Guillermo A. Gonz\'alez 
}


\institute{Oscar M. Pimentel \at
              Grupo de Investigaci\'on en Relatividad y Gravitaci\'on, 
 Escuela de F\'isica, Universidad Industrial de Santander, A. A. 678,
Bucaramanga 680002, Colombia. \\
              \email{oscar.pimentel@correo.uis.edu.co}           
           \and
            F. D. Lora-Clavijo \at
              Grupo de Investigaci\'on en Relatividad y Gravitaci\'on, 
Escuela de F\'isica, Universidad Industrial de Santander, A. A. 678,
Bucaramanga 680002, Colombia. \\
              \email{fadulora@uis.edu.co}
           \and
           Guillermo A. Gonz\'alez \at
              Grupo de Investigaci\'on en Relatividad y Gravitaci\'on, 
Escuela de F\'isica, Universidad Industrial de Santander, A. A. 678,
Bucaramanga 680002, Colombia. \\
              \email{guillermo.gonzalez@saber.uis.edu.co}
}

\date{Received: date / Accepted: date}

\maketitle

\begin{abstract}
Considering the growing interest of the astrophysicist community in the study of dissipative fluids with the aim of getting a more realistic description of the universe, we present in this paper a physical analysis of the energy-momentum tensor of a viscous fluid with heat flux. We introduce the general form of this tensor and, using the approximation of small velocity gradients, we relate the stresses of the fluid with the viscosity coefficients, the shear tensor and the expansion factor. Exploiting these relations, we can write the stresses in terms of the extrinsic curvature of the normal surface to the 4-velocity vector of the fluid, and we can also establish a connection between the perfect fluid and the symmetries of the spacetime. On the other hand, we calculate the energy conditions for a dissipative fluid through contractions of the energy-momentum tensor with the 4-velocity vector of an arbitrary observer. This method is interesting because it allows us to compute the conditions in a reasonably easy way and without considering any approximation or restriction on the energy-momentum tensor.
\end{abstract}

\section{Introduction}
\label{intro}
Perfect fluids are of great interest since they have been used to describe different astrophysical scenarios in the context of the general relativity. In fact, most numerical relativity simulations are carried out considering fluids in which the viscosity and heat conduction vanish \cite{2013rehy.book.....R,2010nure.book.....B}. Nevertheless, sometimes it is important to consider the dissipative effects. This is the case of neutron star oscillations, for which a perfect fluid model may present instabilities, being important to introduce dissipative effects to study whether the instability disappears or not \cite{2003CQGra..20R.105A}. Moreover, the relation between the viscosity and the stability of a star is fundamental for gravitational-wave asteroseismology \cite{2007LRR....10....1A}.  On the other hand, recent efforts to understand the importance of viscous fluids in the context of relativistic heavy ion collision have taken relevance since phenomenological investigations have shown that the bulk viscosity of QCD matter is not small \cite{2015PhRvL.115m2301R}.

Now, if we want to include dissipative terms in the energy-momentum tensor, in order to obtain solutions that describe more realistic astrophysical objects, we can not forget that these solutions need to satisfy some energy conditions that ensure a reasonable physical behavior of the gravitational field source \cite{1984ucp..book.....W,0264-9381-5-10-011}. These conditions are very useful to reduce the number of possible solutions to the Einstein equations to only those with physical meaning \cite{2004sgig.book.....C}, and to prove the occurrence of singularities without requiring the exact expression of the energy-momentum tensor \cite{1973lsss.book.....H}. On the other hand, from the point of view of the numerical simulations, a problem one has to face is that the general relativistic Eulerian hydrodynamics equations may diverge or develop unphysical results once the rest mass density tends to zero; thus a numerical atmosphere has to be implemented to prevent these kind of effects. Commonly, the atmosphere is defined as the minimum value of the rest mass density, that avoids the divergence of the specific enthalpy and possible errors which may be propagated to the other variables. For this reason, the implementation of an atmosphere based on the energy conditions makes the numerical simulations more physically credible.  

Based on the above discussion, it is important to recognize the physics of dissipative processes contained in the energy-momentum tensor of an imperfect fluid; in other words, to understand how is the heat flux vector related to the temperature, and how is the stress tensor related to the viscosity. The first relation is not difficult to find and understand since many books and papers explain it \cite{1973grav.book.....M,2010nure.book.....B,2007LRR....10....1A,Maartens:1996vi}. Nevertheless, the relation between the stresses and the viscosity is not so clear. 
For this reason, the first objective of this work is to relate the stresses of the fluid with the viscosity coefficients and the kinematic quantities of the medium (shear and expansion), under the approximation of small velocity gradients. It allows us to write the main stresses of the energy-momentum tensor in terms of the geometric properties, and also, to relate the isotropy of the fluid with the symmetries of the spacetime. 

The second part of this paper is devoted to determining the energy conditions for a fluid with viscosity and heat flux. These conditions have already been computed by Kolassis, Santos and Tsoubelis in 1988 \cite{0264-9381-5-10-011} by means of the method of eigenvalues and eigenvectors of the energy-momentum tensor. Nevertheless, this method leads to a fourth order characteristic polynomial whose roots are too complicated to be found explicitly, and it requires the introduction of some restrictions. Fortunately, following the method used by E. Poisson \cite{2004rtmb.book.....P} to compute the energy conditions for a viscous fluid, we can find the general energy conditions of a fluid with viscosity and non-zero heat flux vector more easily and in a natural way, ${\it i.e.}$ without dealing with fourth order polynomials or introducing restrictions on the source of gravitational field. With this method, we only need to write the spatial part of the energy-momentum tensor (corresponding to the stress tensor) in an orthonormal basis in which the diagonal elements are the principal stresses of the fluid and the heat flux 4-vector has arbitrary components. This method provides a general procedure for writing down the energy conditions without the assumption of small velocity gradients. 

The layout of this paper is the following: in section \ref{sec:EMT}, we introduce the energy-momentum tensor for a viscous fluid with non-zero heat flux, and we analyze how the stresses are related to the kinematic properties (shear and expansion) and the viscosity coefficients, in the approximation of small velocity gradients. Additionally, we present some interesting equations that establish a relationship between the stresses and the geometric properties of the spacetime. In section \ref{sec:EC}, we compute the energy conditions for a viscous fluid with heat flux by means of the physical quantities (energy density, energy flux density, etc.) as measured by an arbitrary observer. This procedure is interesting because of its simplicity and because it is free from restrictions. Finally, we present our main concluding remarks in section \ref{sec:conclusions}. We will work with the signature $(-,+,+,+)$, and with units in which the gravitational constant $G$ and the speed of light $c$ are equal to one.

\section{The Energy-Momentum Tensor}
\label{sec:EMT}
In general relativity, the sources of the gravitational field are the energy and the momentum of the particles forming the astrophysical object. This information is given by the energy-momentum tensor $T^{\alpha\beta}$, which represents the flux of the $\alpha$-component of the momentum across a surface where the coordinate $x^{\beta}$ is constant \cite{1985fcgr.book.....S}. In this way, the $T^{00}$ component of the energy-momentum tensor is the energy flux across a surface of $t$ constant, ${\it i.e.}$ the energy density. Following a similar reasoning, $T^{0i}$ is the energy flux across a surface of $x^{i}$ constant. Now, if the frame of reference moves with the fluid (comoving reference frame), then the particles will have zero velocity and the energy will only flow via heat conduction, so $T^{0i}$ corresponds to the heat flux in the $i$ direction. Finally, $T^{ij}$ is the flux of the $i$ momentum across the surface of $x^{j}$ constant, hence, it represents the stress in the $i$ direction on the surfaces of $x^{j}$ constant. The components $T^{0i}$ and $T^{ij}$ are related to dissipative phenomena like thermal radiation and viscosity as we will see later.
	
We will consider a general anisotropic fluid with nonzero heat flux vector, whose energy--momentum tensor can be written as follows,
\begin{equation}\label{primertab}
T^{\alpha \beta}=\rho u^{\alpha} u^{\beta} + q^{\alpha} u^{\beta} +
q^{\beta} u^{\alpha} + S^{\alpha \beta}\, ,
\end{equation}
where $\rho$ is the energy density as measured by a ``comoving'' observer
with velocity $u^{\alpha}$, which satisfies $u^{\alpha}u_{\alpha}=-1$ with $u^0
> 0$, $q^{\alpha}$ is the spacelike ``heat'' flux vector such that 
$q^{\alpha}u_{\alpha}=0$, and $S^{\alpha \beta}$ is the stress tensor. The eigenvectors of the stress tensor define the directions of the principal stresses as a base in which $S^{\alpha\beta}$ is diagonal. By means of a Gram--Schmidt process the base becomes orthonormal in such a way that
\begin{eqnarray}\label{baseespacial}
&&x^{\alpha}y_{\alpha}=0\, ,\qquad x^{\alpha}x_{\alpha}=1\, ,\qquad x^{\alpha}u_{\alpha}=0\, ,\\
&&x^{\alpha}z_{\alpha}=0\, ,\qquad y^{\alpha}y_{\alpha}=1\, ,\qquad y^{\alpha}u_{\alpha}=0\, ,\\
&&y^{\alpha}z_{\alpha}=0\, ,\qquad z^{\alpha}z_{\alpha}=1\, ,\qquad
z^{\alpha}u_{\alpha}=0\, ,
\end{eqnarray}
and the eigenvalues equations are
\begin{equation}
S^{\alpha \beta} x_\beta = p_1 x^\alpha, ~ ~ S^{\alpha \beta}
y_\beta = p_2 y^\alpha, ~ ~ S^{\alpha \beta} z_\beta = p_3
z^\alpha ,
\end{equation}
where we have defined $p_i$, with $i=1, 2, 3,$ as the principal stresses.

Thus, the energy-momentum tensor in the ``comoving'' tetrad
\begin{equation}
e^{\alpha}_{(\mu)}=\{e^{\alpha}_{(0)},e^{\alpha}_{(1)},e^{\alpha}_{(2)},e^{\alpha}_{(3)}\}=\{ u^\alpha , x^\alpha , y^\alpha , z^\alpha
\},
\end{equation}
is given by (\ref{primertab}) with
\begin{equation}\label{stresstensor}
S^{\alpha \beta}=p_1 x^\alpha x^\beta + p_2 y^\alpha y^\beta + p_3
z^\alpha z^\beta,
\end{equation}
and,
\begin{equation}\label{fluxvector}
q^{\alpha}=q_1 x^\alpha + q_2 y^\alpha + q_3 z^\alpha.
\end{equation}
It is worth mentioning that the heat flux vector has arbitrary components because its direction does not
coincide necessarily with one of the principal stresses. This vector
satisfies the relations
\begin{equation}
q^\alpha u_\alpha = 0 , \hspace{2mm} q^\alpha q_\alpha = q^2 \geq 0,
\end{equation}
where $q = \sqrt{q_1^2 + q_2^2 + q_3^2}$.

When the fluid is at rest, the tangential stresses are zero and the normal stress does not depend on the direction of the normal vector \cite{2000ifd..book.....B}. In this case $p_{1}=p_{2}=p_{3}=p$, and the stress tensor (\ref{stresstensor}) becomes $S^{\alpha\beta}=p(x^{\alpha}x^{\beta}+y^{\alpha}y^{\beta}+z^{\alpha}z^{\beta})$. We identify the metric of the spatial surface (normal to $u^{\alpha}$) $h^{\alpha\beta}=x^{\alpha}x^{\beta}+y^{\alpha}y^{\beta}+z^{\alpha}z^{\beta}$, which is related to the metric of the spacetime through the equation
\begin{equation}\label{inducedmetric}
h_{\alpha\beta}=g_{\alpha\beta}+u_{\alpha}u_{\beta}.
\end{equation} 
In this way, the stress tensor can be written as
\begin{equation}\label{staticstress}
S^{\alpha\beta}=ph^{\alpha\beta},
\end{equation}
where $p$ is the local equilibrium pressure of the fluid or static pressure.

When the fluid is moving, the situation is quite different, the tangential stresses are different from zero, and the normal stress depends on the spatial direction, ${\it i.e.}$ $p_{1}\neq p_{2}\neq p_{3}$ in general. As in the static fluid, it is interesting to have a physical quantity, which will be denoted as $\hat{p}$ to differentiate it from the local equilibrium pressure $p$, that measures the intensity of the ``squeezing" in a point of the fluid. Therefore, we take this quantity to be the average value of the normal component of the stress over the surface of a small sphere centered at the point of interest. It can be shown that this quantity is \cite{2000ifd..book.....B}
\begin{equation}\label{isotropic}
\hat{p}=\frac{S}{3}=\frac{p_{1}+p_{2}+p_{3}}{3},
\end{equation}
where $S$ is the trace of $S^{\alpha\beta}$. This local average stress is usually called the isotropic pressure, it is invariant under rotations of the spatial axes, and its interpretation in a moving fluid is similar to that of $p$. Now, it is convenient to write the stress tensor as 
\begin{equation}\label{stressmoving}
S^{\alpha\beta}=\hat{p}h^{\alpha\beta}+\Pi^{\alpha\beta},
\end{equation}
where $\hat{p}h^{\alpha\beta}$ is an isotropic term, and $\Pi^{\alpha\beta}$ is a traceless tensor that contains information about the tangential stresses (anisotropic term). Substituting (\ref{stressmoving}) in the energy-momentum tensor (\ref{primertab}), we obtain
\begin{equation}\label{tercerotab}
T^{\alpha \beta}=\rho u^{\alpha} u^{\beta} + \hat{p} h^{\alpha \beta}
+ q^{\alpha} u^{\beta} + q^{\beta} u^{\alpha} + \Pi^{\alpha \beta}.
\end{equation}

On the other hand, according to Landau and Lifshitz \cite{1959flme.book.....L}, processes of internal friction (viscosity) occur when there is a relative velocity between two adjacent elements of the fluid. Now, if the velocity gradients are small, we can consider the stress tensor depending only on the first derivatives of the velocity, as a linear combination of them. Using this first order approximation, Eckart proposed in 1940 a relativistic theory for dissipative fluids \cite{1940PhRv...58..919E}, which is widely used nowadays, despite its causality and stability problems \cite{Maartens:1996vi}. In the Eckart theory, the energy-momentum tensor is written in terms of the viscosity coefficients as \cite{1973grav.book.....M}
\begin{equation}\label{emt2}
T^{\alpha \beta}=\rho u^{\alpha} u^{\beta} + q^{\alpha} u^{\beta} +
q^{\beta} u^{\alpha} + (p - \zeta \Theta) h^{\alpha \beta} - 2 \eta \sigma^{\alpha \beta} \, ,
\end{equation}
where $\zeta > 0$ and $\eta > 0$  are the bulk and shear viscosity, respectively, 
$\Theta = u^{\alpha}_{; \alpha}$ is the expansion, and $\sigma^{\alpha \beta}$ is the shear tensor,
\begin{equation}\label{shear}
\sigma^{\alpha \beta}=\frac{1}{2} \left(u^{\alpha}_{; \mu}h^{\mu \beta} 
+ u^{\beta}_{; \mu}h^{\mu \alpha} \right) - \frac{1}{3}\Theta h^{\alpha \beta},
\end{equation}
which is symmetric and traceless. Physically, the expansion measures the rate of change of the volume of a fluid element per unit volume, and the shear tensor measures the shearing deformation of a fluid element.

As we can see, it is possible to write down  the expression (\ref{emt2}) as
the sum of three components
\begin{equation}\label{emt3}
T^{\alpha \beta} = T_{pf}^{\alpha \beta} + T_{heat}^{\alpha \beta} + T_{visc}^{\alpha \beta}, 
\end{equation}
where
\begin{equation}\label{emt3}
\nonumber T_{pf}^{\alpha \beta} =  \rho u^{\alpha} u^{\beta} + p h^{\alpha \beta}, ~~~ 
\nonumber T_{heat}^{\alpha \beta} = q^{\alpha} u^{\beta} + q^{\beta} u^{\alpha},  ~~~
\nonumber T_{visc}^{\alpha \beta} = - \zeta \Theta h^{\alpha \beta} - 2 \eta \sigma^{\alpha \beta}, 
\end{equation}
are the perfect fluid, heat flux and viscosity energy momentum tensors, respectively.

On the other hand, according to the non-relativistic Fourier's law of heat conduction, the heat flows when the temperature changes between two spatial points in the fluid. When the temperature gradient $\nabla T$ is small, the non-relativistic heat flux can be written, in an approximate manner, as $-\kappa\nabla T$, where the constant $\kappa$ is the thermal conductivity and the minus sign states that the heat flows from a point of higher temperature to another point where $T$ is lower. Generalizing the non-relativistic Fourier's law of heat conduction for a relativistic fluid, and taking into account the flow of heat due to the accelerated matter (relativistic effect), the heat flux vector can be written as \cite{1940PhRv...58..919E}
\begin{equation}\label{heatflux}
q^{\alpha}=-\kappa h^{\alpha\beta}(T_{,\beta}+Ta_{\beta}),
\end{equation}
where $a_{\beta}=u^{\alpha}u_{\beta;\alpha}$ is the acceleration of the fluid, so the second term in the last equation is an isothermal heat flux due to the inertia of the energy.

Now, it is reasonable to think that the isotropic pressure and the main stresses in (\ref{primertab}) and (\ref{tercerotab}) should be related to the viscosity coefficients, the shear tensor, and the expansion in (\ref{emt2}). To obtain such relations, we project (\ref{tercerotab}) and (\ref{emt2}) on the induced spatial metric, so we find that the isotropic pressure, $\hat{p} = \frac{1}{3}h_{\alpha \beta}T^{\alpha \beta}$, is given by 
\begin{equation}\label{relat1}
\hat{p} = p - \zeta \Theta,
\end{equation}
so $\zeta \Theta$ is a bulk viscous pressure, which measures the deviation of $\hat{p}$ from the local equilibrium pressure $p$. Using the last expression and comparing (\ref{tercerotab}) and (\ref{emt2}), it is easy to see that $\Pi^{\alpha \beta}$ can be written as
\begin{equation}\label{relat2}
\Pi^{\alpha \beta} =  S^{\alpha \beta}- \hat{p} h^{\alpha \beta} = -2 \eta \sigma^{\alpha \beta}.
\end{equation}
This expression is very interesting because we can project its two indices with the spatial vectors of the comoving tetrad, ($x_{\alpha},y_{\alpha}, z_{\alpha}$), in order to get the principal stresses $p_1, p_2$ and $p_3$ in terms of $\eta$ and $\sigma^{\alpha\beta}$, such that
\begin{equation}\label{stresses}
p_1 = \hat{p} - 2 \eta \sigma^{\alpha \beta}x_{\alpha} x_{\beta},  ~~~~ 
p_2 = \hat{p} - 2 \eta \sigma^{\alpha \beta}y_{\alpha} y_{\beta},  ~~~~
p_3 = \hat{p} - 2 \eta \sigma^{\alpha \beta}z_{\alpha} z_{\beta}. 
\end{equation}

The expressions (\ref{relat1}) and (\ref{stresses}) allow us to analyze the relation between the stresses and the dynamic properties of the fluid. We can distinguish the following four cases:
\begin{itemize}
\item{\bf CASE 1:} If the properties of fluid are such that the bulk and shear viscosity are zero, ${\it i.e.}$ $\zeta=0$ and $\eta=0$ (no viscosity), then $p_{1}=p_{2}=p_{3}=\hat{p}=p$. We can say that the stress in the fluid is isotropic.

\item{\bf CASE 2:} If the fluid velocity is zero (static case), the expansion and the shear vanishes and the fluid does not dissipate energy via viscosity. This implies again that $p_{1}=p_{2}=p_{3}=\hat{p}=p$ and the fluid becomes isotropic.

\item{\bf CASE 3:} If $\eta=0$ and $\zeta\neq 0$ then $p_{1}=p_{2}=p_{3}=\hat{p}$, and $\hat{p}=p+\zeta\Theta$. In this case, the stress of the fluid is isotropic, but $\hat{p}$ is not just the equilibrium pressure, there is an additional isotropic contribution to the energy-momentum tensor of the form $\zeta\Theta h^{\alpha\beta}$ due to the bulk viscosity \cite{2010CQGra..27t5012C}.

\item{\bf CASE 4:} If $\eta\neq 0$ and $\zeta=0$ then the bulk viscosity contribution to the isotropic pressure vanishes, so $\hat{p}=p$. Nevertheless, the stress of the fluid is anisotropic since $\sigma^{\alpha \beta}x_{\alpha} x_{\beta}\neq \sigma^{\alpha \beta}y_{\alpha} y_{\beta}\neq \sigma^{\alpha \beta}z_{\alpha} z_{\beta}\neq 0 $ in general. It means that the energy is dissipating due to shearing deformations.
\end{itemize}
If $q^{\alpha}=0$, then the cases 1 and 2 will describe a perfect fluid (no energy dissipation).

On the other hand, there is an alternative way for writing (\ref{stresses}) in terms of the extrinsic curvature of the normal surface to $u^{\alpha}$, which is defined as
\begin{equation}
K_{ab}=u_{\alpha;\beta}e_{(a)}^{\alpha}e_{(b)}^{\beta},
\label{extrincic_curvature}
\end{equation}
where $a=1,2,3$. This curvature provides information on how much the 4-velocity direction changes from one point to another in the spatial surface \cite{2010nure.book.....B}. To do it, we start by projecting (\ref{shear}) along the spatial components of the tetrad, such that
\begin{eqnarray}
\nonumber \sigma^{\alpha\beta}x_{\alpha}x_{\beta}&=&u_{\alpha;\beta}x^{\alpha}x^{\beta}-\frac{\Theta}{3},\\ 
\sigma^{\alpha\beta}y_{\alpha}y_{\beta}&=&u_{\alpha;\beta}y^{\alpha}y^{\beta}-\frac{\Theta}{3}, \label{first_proyection}\\
\nonumber \sigma^{\alpha\beta}z_{\alpha}z_{\beta}&=&u_{\alpha;\beta}z^{\alpha}z^{\beta}-\frac{\Theta}{3};
\end{eqnarray}
nevertheless, according to (\ref{extrincic_curvature}), the double contraction of $u_{\alpha;\beta}$ with the spatial vectors $x^{\alpha}$, $y^{\alpha}$ and $z^{\alpha}$ are the components $K_{11}$, $K_{22}$ and $K_{33}$ of the the extrinsic curvature respectively. So, we can rewrite (\ref{first_proyection}) as
\begin{eqnarray}
\nonumber \sigma^{\alpha\beta}x_{\alpha}x_{\beta}&=&K_{11}-\frac{\Theta}{3},\\
\sigma^{\alpha\beta}y_{\alpha}y_{\beta}&=&K_{22}-\frac{\Theta}{3},\label{first_curvature}
\\
\nonumber \sigma^{\alpha\beta}z_{\alpha}z_{\beta}&=&K_{33}-\frac{\Theta}{3},
\end{eqnarray}
in such a way that (\ref{stresses}) becomes
\begin{eqnarray}
\nonumber p_1 &=& \hat{p} - 2 \eta \left(K_{11}-\frac{\Theta}{3} \right),  \\
p_2 &=& \hat{p} - 2 \eta \left(K_{22}-\frac{\Theta}{3} \right), \label{stresses_curvature}
\\
\nonumber p_3 &=& \hat{p} - 2 \eta \left(K_{33}-\frac{\Theta}{3} \right). 
\end{eqnarray}
According to the last expressions, we distinguish the next case:
\begin{itemize}
\item{\bf CASE 5:} When $K_{11}=K_{22}=K_{33}=\Theta/3$, the stresses are such that $p_{1}=p_{2}=p_{3}=\hat{p}$ and $\hat{p}=p+\zeta\Theta$; therefore the fluid becomes isotropic and the energy is dissipated through bulk viscosity. So, we can also study the energy dissipation due to the bulk viscosity when the diagonal terms of the extrinsic curvature are equal to $\Theta/3$.

\end{itemize}
Even though in the cases 3 and 5 the fluids have isotropic stresses and the bulk viscosity is different from zero, these are physically different since in the case 3 we made an assumption about the viscosity coefficients, which does not say anything about the state of motion of the fluid, while in the case 5 the assumption constrains the movement of the fluid.

Finally, there is another interesting alternative for writing (\ref{stresses}) in order to analyze the stresses of the fluid, but now in terms of the symmetries of the spacetime. To do this, we start writing the first terms in (\ref{first_proyection}) as
\begin{eqnarray}
\nonumber u_{\alpha;\beta}x^{\alpha}x^{\beta}&=&\frac{1}{2}\left(u_{\alpha;\beta}+u_{\beta;\alpha}\right)x^{\alpha}x^{\beta},\\
u_{\alpha;\beta}y^{\alpha}y^{\beta}&=&\frac{1}{2}\left(u_{\alpha;\beta}+u_{\beta;\alpha}\right)y^{\alpha}y^{\beta},\label{final}
\\
\nonumber u_{\alpha;\beta}z^{\alpha}z^{\beta}&=&\frac{1}{2}\left(u_{\alpha;\beta}+u_{\beta;\alpha}\right)z^{\alpha}z^{\beta}.
\end{eqnarray}
At this point, we need to introduce the well known {\em Lie derivative} \cite{1984ucp..book.....W}, which describes the change of an arbitrary tensor, say $A_{\alpha\beta}$, in the direction of a vector $\vec{\xi}$, and its expression is as follow
\begin{equation}
\pounds_{\vec{\xi}}A_{\alpha\beta}=\xi^{\mu}A_{\alpha\beta;\mu}+A_{\mu\beta}\xi^{\mu}_{\;\; ;\alpha}+A_{\alpha\mu}\xi^{\mu}_{\;\; ;\beta}.
\label{lie}
\end{equation}
This derivative is very interesting because, when $A_{\alpha\beta}$ is the metric tensor of the spacetime, the last expression reduces to
\begin{equation}
\pounds_{\vec{\xi}}g_{\alpha\beta}=\xi_{\alpha;\beta}+\xi_{\beta;\alpha}.
\label{lie1}
\end{equation}
Moreover, if $\vec{\xi}$ defines a direction in which the spacetime does not change (symmetry direction), then
\begin{equation}
\pounds_{\vec{\xi}}g_{\alpha\beta}=0,
\label{lie3}
\end{equation}
and $\vec{\xi}$ becomes in a Killing vector. 

Now, changing $\vec{\xi}$ by $\vec{u}$ in (\ref{lie1}), we can transform (\ref{final}) into
\begin{eqnarray}
\nonumber u_{\alpha;\beta}x^{\alpha}x^{\beta}&=&\frac{1}{2}\left(\pounds_{\vec{u}}g_{\alpha\beta}\right)x^{\alpha}x^{\beta},\\
u_{\alpha;\beta}y^{\alpha}y^{\beta}&=&\frac{1}{2}\left(\pounds_{\vec{u}}g_{\alpha\beta}\right)y^{\alpha}y^{\beta},\label{step1}
\\
\nonumber u_{\alpha;\beta}z^{\alpha}z^{\beta}&=&\frac{1}{2}\left(\pounds_{\vec{u}}g_{\alpha\beta}\right)z^{\alpha}z^{\beta},
\end{eqnarray}
such that (\ref{first_proyection}) takes the following form \cite{1985JMP....26.2881M}
\begin{eqnarray}
\nonumber \sigma^{\alpha\beta}x_{\alpha}x_{\beta}&=&\frac{1}{2}\left(\pounds_{\vec{u}}g_{\alpha\beta}\right)x^{\alpha}x^{\beta}-\frac{\Theta}{3},\\
\sigma^{\alpha\beta}y_{\alpha}y_{\beta}&=&\frac{1}{2}\left(\pounds_{\vec{u}}g_{\alpha\beta}\right)y^{\alpha}y^{\beta}-\frac{\Theta}{3},\label{auxiliar}
\\
\nonumber \sigma^{\alpha\beta}z_{\alpha}z_{\beta}&=&\frac{1}{2}\left(\pounds_{\vec{u}}g_{\alpha\beta}\right)z^{\alpha}z^{\beta}-\frac{\Theta}{3};
\end{eqnarray}
in this way, with (\ref{auxiliar}) we can write (\ref{stresses}) as
\begin{eqnarray}
\nonumber p_1 &=& \hat{p} - 2 \eta \left[\frac{1}{2}\left(\pounds_{\vec{u}}g_{\alpha\beta}\right)x^{\alpha}x^{\beta}-\frac{\Theta}{3} \right], \\
p_2 &=& \hat{p} - 2 \eta \left[\frac{1}{2}\left(\pounds_{\vec{u}}g_{\alpha\beta}\right)y^{\alpha}y^{\beta}-\frac{\Theta}{3} \right], \label{stresses_lie1}
\\
\nonumber p_3 &=& \hat{p} - 2 \eta \left[\frac{1}{2}\left(\pounds_{\vec{u}}g_{\alpha\beta}\right)z^{\alpha}z^{\beta}-\frac{\Theta}{3} \right]. 
\end{eqnarray}
Additionally, the expansion factor can be expressed in terms of the Lie derivative as \cite{2004rtmb.book.....P}
\begin{equation}
\Theta=\frac{1}{2}h^{\alpha\beta}\pounds_{\vec{u}}g_{\alpha\beta},
\label{expansion_lie}
\end{equation}
so (\ref{stresses_lie1}) reduces to
\begin{eqnarray}
\nonumber p_1 &=& \hat{p} - \eta \left(x^{\alpha}x^{\beta}-\frac{1}{3}h^{\alpha\beta}\right)\pounds_{\vec{u}}g_{\alpha\beta}, \\
p_2 &=& \hat{p} - \eta \left(y^{\alpha}y^{\beta}-\frac{1}{3}h^{\alpha\beta}\right)\pounds_{\vec{u}}g_{\alpha\beta}, \label{first_curvature}
\\
\nonumber p_3 &=& \hat{p} - \eta \left(z^{\alpha}z^{\beta}-\frac{1}{3}h^{\alpha\beta}\right)\pounds_{\vec{u}}g_{\alpha\beta},
\end{eqnarray}
and (\ref{relat1}) turns into
\begin{equation}
\hat{p}=p-\frac{1}{2}\zeta h^{\alpha\beta}\pounds_{\vec{u}}g_{\alpha\beta};
\label{isotropic_lie}
\end{equation}
therefore, we can state that:
\begin{itemize}
\item {\bf CASE 6:} If the 4-velocity vector of the fluid, $u^{\alpha}$, is a linear combination of the Killing vectors of the spacetime, ${\it i.e.}$ if the fluid moves in world lines in which the properties of the spacetime does not change, then $\pounds_{\vec{u}}g_{\alpha\beta}=0$ and the fluid becomes isotropic with $p_{1}=p_{2}=p_{3}=\hat{p}=p$. If in addition $q^{\alpha}=0$, then the fluid will be perfect. 
\end{itemize} 
This case is very interesting because it establishes a connection between the spacetime symmetries, the movement of the fluid, and the dissipation phenomena due to viscosity. 

In this section, we have analyzed the main stresses of a dissipative  fluid with heat flux in terms of viscosity 
coefficients,  the shear tensor and the expansion, under the approximation of small velocity gradients (Eckart framework). 
However, in order to guarantee the existence of solutions to the Einstein equations in the presence of a
realistic viscous fluid, the source must obeys the energy conditions. Such conditions are computed in the next section 
for a fluid with different main stresses and a non-zero heat flux vector, without any assumptions concerning to the velocity gradients.

\section{The Energy Conditions}
\label{sec:EC}
Any realistic gravitational system, described with the Einstein field equations
\begin{equation}
R_{\alpha \beta} - \frac{1}{2} g_{\alpha \beta} R = 8 \pi T_{\alpha \beta}, \label{eq:Einstein}
\end{equation}
where $R_{\alpha \beta}$ is the Ricci tensor and $R$ is the scalar of  curvature, needs to satisfy some energy conditions that ensure the reasonable physical behavior of the matter and its gravitational field. On the other hand, there are infinite possibilities for choosing the energy-momentum tensor to get a solution of (\ref{eq:Einstein}), so we need to impose the energy conditions to reduce them to only those with physical meaning \cite{2004sgig.book.....C}.

Now, in order to find the energy conditions for a relativistic dissipative fluid, we need an observer that measures the thermodynamic quantities (energy density, stresses, energy fluxes, and so on); therefore, we consider an arbitrary observer defined by its 4-velocity vector  $W^{\alpha}$, which satisfies the following conditions: $W^\alpha W_\alpha = - 1$  and $W^0 >	0$, that is, the vector is timelike and also future oriented, respectively. In terms of the comoving tetrad, we can write 
\begin{equation}\label{arbitrary}
W^\alpha = W^{(\mu)} e_{(\mu)}^\alpha = \gamma u^\alpha + A_1 x^\alpha +
A_2 y^\alpha + A_3 z^\alpha ,
\end{equation}
where $\gamma$ and $A_{i}$ are the components of the $W^{\alpha}$ in the comoving tetrad. Since the 4-velocity is timelike and unitary, its components have to satisfy the following relation 
\begin{equation}
\gamma = \sqrt{1 + A_1^2 + A_2^2 + A_3^2} \geq 1. \label{eq:gam}
\end{equation}
The vector $A^\alpha = A_1 x^\alpha + A_2 y^\alpha + A_3 z^\alpha$
is the spatial part of $W^\alpha$, such that
\begin{eqnarray}
\nonumber A_\alpha u^\alpha &=& 0 ,~~~~~~ A_\alpha A^\alpha = A^2 \geq 0 , \\ 
A &=& \sqrt{A_1^2 + A_2^2 + A_3^2} \geq 0 .
\end{eqnarray}
From the last two relations, we can see that $\gamma = \sqrt{1 + A^2}$,
and therefore, when $A$ goes to infinity, the following condition 
\begin{equation}
\lim_{A \to \infty} \gamma = A ,
\label{eq:lim2}
\end{equation}
has to be satisfied, so that $\gamma \geq A$. 

Now, we can compute the thermodynamics quantities, as measured by the arbitrary observer, by contracting the energy-momentum tensor (\ref{primertab}) with the 4-velocity vector $W^{\alpha}$. So we can define the quantities
\begin{eqnarray}\label{eq:epsilon}
\epsilon &=& T_{\alpha\beta} W^\alpha W^\beta , \\ 
S_\alpha &=& - T_{\alpha\beta} W^\beta,
\end{eqnarray}
where $\epsilon$ is the energy density and $S_a$ is the energy flux density vector, both measured by $W^\alpha$. In terms of the above quantities we formulate the energy conditions as:

\begin{enumerate}

\item {\it Weak Energy Condition (WEC)}. The energy density as measured by any arbitrary observer can not be negative:
\begin{equation}
    \epsilon \geq 0 .
\end{equation}

\item {\it Strong Energy Condition (SEC)}. Any arbitrary timelike or null congruence must be convergent, thus the gravitational field will be attractive:
\begin{equation}
\label{SenergyC}
    R_{\alpha\beta} W^\alpha W^\beta \geq 0.
\end{equation}
Nevertheless, from expression (\ref{eq:Einstein}) the Ricci tensor, $R_{\alpha\beta}$, can be written in terms of $T_{\alpha\beta}$ as
\begin{equation}
\label{riccienrgy}
R_{\alpha\beta}=8\pi\left(T_{\alpha\beta}-\frac{1}{2}Tg_{\alpha\beta} \right),
\end{equation}
where $T = -R/8\pi$; then, the relation (\ref{SenergyC}) can be also written as
\begin{equation}\label{eq:mu}
    \mu = \epsilon + \frac{T}{2} \geq 0 , 
\end{equation}
where $T = g^{\alpha\beta} T_{\alpha\beta}$, the trace of $T_{\alpha\beta}$, is given by
\begin{equation}
T = - \rho + p_1 + p_2 + p_3 = - \rho + 3 \hat{p}.
\end{equation}

\item {\it Dominant Energy Condition (DEC)}. The energy flux density vector, as measured by any arbitrary observer, must be a future oriented timelike or null vector:
\begin{eqnarray}
S_\alpha S^\alpha &\leq& 0 , \label{do1} \\
S^0 &>& 0 \label{do2}.
\end{eqnarray}

\end{enumerate}

In this work, and in order to illustrate the procedure, we present two cases. The first one corresponds to the case when the observer is moving with the fluid and the second one  is  the general case in which the observer moves arbitrarily. 
In the first case, the energy conditions at the comoving tetrad take a very simple form and they are obtained when $W^\alpha = u^\alpha$; that is, by taking $A_i = 0$ and $\gamma=1$. In this case we have that the energy conditions reduce to
\begin{eqnarray}
WEC &:& \rho \geq 0  , \\
SEC &:& \rho + 3\hat{p} \geq 0 , \\
DEC &:& \rho \geq q .
\end{eqnarray}
However, in the second case for an arbitrary non-comoving observer, the procedure to get the energy conditions is more complicated and for this reason,  we will analyze each one, with more detail, in the following subsections.

\subsection{Weak Energy Condition}

From (\ref{eq:epsilon}), we can write the energy density $\epsilon$ in terms of the quantities measured by the comoving observer as
\begin{equation}
\epsilon = \rho \gamma^2 + \sum_{i=1}^{3} p_i A_i^2- 2 \gamma ( {\bf
q} \cdot {\bf A} ) ,
\end {equation}
where
\begin{equation}
{\bf q} \cdot {\bf A} = \sum_{i = 1}^3 q_i A_i ,
\end{equation}
is the usual scalar product in the euclidean space. Now, by using
(\ref{eq:gam}), we can write
\begin{equation}
\epsilon = \rho + \sum_{i=1}^{3}(\rho + p_i) A_i^2 - 2 \gamma ( {\bf
q} \cdot {\bf A} ).
\end {equation}
On the other hand, it is clear that
\begin{equation}
- q A \leq {\bf q} \cdot {\bf A} \leq q A, \label{eq:dot}
\end{equation}
therefore, $\epsilon$ will take the lowest value when ${\bf q}\cdot {\bf A}=qA$, so we can state that
\begin{equation}
\epsilon \geq \rho + \sum_{i=1}^{3}(\rho + p_i) A_i^2 - 2 q A
\gamma.
\end{equation}
Finally, according to (\ref{eq:lim2}), we can use $A=\gamma$ to ensure, again, the lowest value for $\epsilon$, in such a way that
\begin{equation}
\epsilon \geq \rho + \sum_{i=1}^{3}(\rho + p_i) A_i^2 - 2 q \gamma^2,
\end{equation}
then, it can be written as
\begin{equation}
\epsilon \geq (\rho - 2 q) + \sum_{i=1}^{3}(\rho + p_i-2 q) A_i^2,
\end{equation}
where we have used (\ref{eq:gam}). Therefore, $\epsilon \geq 0$ for all values of $A_{i}$ if the function 
$(\rho - 2 q) + \sum_{i=1}^{3}(\rho + p_i-2 q) A_i^2$ has a positive minimum with respect to the variables $(A_1,A_2,A_3)$. It happens if and only if the eigenvalues associated with  the quadratic form 
$\sum_{i=1}^{3}(\rho + p_i-2 q) A_i^2$ are positive and $(\rho - 2 q)$ is positive.  Thus, the necessary and sufficient conditions in order to have $\epsilon \geq 0$, in agreement with the weak energy condition, are 
\begin{eqnarray}
\rho &\geq& 2 q , \\
\rho + p_i &\geq& 2 q ,
\end{eqnarray}
with $i=1,2,3$.
\subsection{Strong Energy Condition}

By using the values of the energy density $\epsilon$ and the trace of the energy-momentum tensor $T$, we have that the expression (\ref{eq:mu}) takes the form
\begin{equation}
\mu = \frac{1}{2}\left(\rho + \sum_{i=1}^{3} p_i\right) +
\sum_{i=1}^{3}(\rho + p_i) A_i^2 - 2 \gamma ( {\bf q} \cdot {\bf A}
).
\end {equation}
Now, following a similar procedure as the one used for the weak energy condition, ${\it i.e.}$ by using (\ref{eq:gam}), (\ref{eq:lim2}) and (\ref{eq:dot}), we can write the last expression as
\begin{equation}
\mu \geq \frac{\rho + 3 p - 4 q}{2} + \sum_{i=1}^{3}(\rho + p_i-2 q)
A_i^2,
\end{equation}
so that, in a similar way as in the weak energy condition, the necessary and sufficient conditions to have
$\mu \geq 0$, in order to fulfill the strong energy condition, are
\begin{eqnarray}
&& \rho + 3 \hat{p} \geq 4 q , \\
&& \rho + p_i \geq 2 q ,
\end{eqnarray}
where $i=1,2,3$.

\subsection{Dominant Energy Condition}

With the aim of finding the dominant energy condition in terms of the energy density $\rho$, the heat flux vector $q^{\alpha}$, and the principal stresses $p_{1}$, $p_{2}$ and $p_{3}$, we start by writing the conditions (\ref{do1}) and (\ref{do2}) in the comoving reference frame, where the metric tensor reduces to that of Minkowski and the calculations are easier. Now, since $S_{\alpha}$ can be taken as a linear combination of the basis vectors $e^{(\mu)}_{\alpha}$ as $S_{\alpha}=S_{(\mu)}e_{\alpha}^{(\mu)}$, we can obtain the components of the energy flux density in the comoving tetrad by multiplying this expression by $e_{\alpha}^{(\mu)}$ and considering the orthonormal properties of the tetrad to get
\begin{equation}
S_{(\mu)} = e_{(\mu)}^{\alpha} S_{\alpha}.
\end{equation}
Additionally, the contravariant components are obtained with the Minkowski metric via $S^{(\mu)} = \eta^{(\mu)(\nu)} S_{(\nu)}$, such that
\begin{equation}\label{s}
S^{(0)}=\rho \gamma - {\bf q} \cdot {\bf A} , \quad S^{(i)}=q_i
\gamma - p_i A_i,
\end{equation}
with $i=1,2,3.$

Now, the dominant energy condition demands the energy flux vector $S^{\alpha}$ to be future oriented; this requirement is equivalent to $S^{(0)} > 0$. However, since
\begin{equation}
S^{(0)} = \rho \gamma - {\bf q} \cdot {\bf A} > (\rho - q) \gamma,
\end{equation}
then the necessary and sufficient condition to have
$S^{(0)} > 0$ is that
\begin{equation}
\rho > q. \label{futur}
\end{equation}
Moreover, the term $S^{\alpha}S_{\alpha}$ in (\ref{do1}) is invariant under coordinate transformations, so we can state that 
\begin{equation}
S_{\alpha} S^{\alpha} = S_{(\mu)} S^{(\mu)} = - (S^{(0)})^2 + (S^{(1)})^2 +
(S^{(2)})^2 + (S^{(3)})^2.
\end{equation}
Thus, if we define
\begin{equation}\label{ns}
N(S) = (S^{(0)})^2 - (S^{(1)})^2 - (S^{(2)})^2 - (S^{(3)})^2,
\end{equation}
then $S_{\alpha} S^{\alpha} = -N(S)$, and the dominant energy condition requires that $N(S)\geq 0$. On the other hand, with (\ref{s}) we can write $N(S)$ as 
\begin{equation}
N(S)=(\rho \gamma - {\bf q} \cdot {\bf A})^2 - \sum_{i=1}^{3}(q_i
\gamma - p_i A_i)^2.
\end{equation}
By expanding this last expression and using the value of
$\gamma$, we obtain
\begin{eqnarray}
N(S)&=&(\rho^2 - q^2) + \sum_{i=1}^{3}(\rho^2 - q^2-p_i^2) A_i^2+
({\bf q} \cdot {\bf A})^2 \nonumber \\&+& 2 \gamma
\left[\sum_{i=1}^{3}q_i p_i A_i - \rho ( {\bf q} \cdot {\bf
A})\right],
\end{eqnarray}
and then, using the condition (\ref{futur}) and the fact that $q \geq 0$, we obtain
\begin{eqnarray}
N(S)&\geq& (\rho^2 - q^2 - 2\rho q) + \sum_{i=1}^{3}(\rho^2 - q^2-
2\rho q - p_i^2) A_i^2\nonumber \\&+& 2 \gamma \sum_{i=1}^{3}q_i p_i
A_i,
\end{eqnarray}
where we have used ${\bf q} \cdot {\bf A} \leq q A \leq q \gamma$
and dropped the term $({\bf q} \cdot {\bf A})^2$ because it is always
positive.
Finally, as $- q \leq q_i \leq q$ and $- A \leq A_i \leq A$, is easy
to see that
\begin{equation}
- q \gamma \leq - q A \leq q_i A_i \leq q A \leq q \gamma,
\end{equation}
and therefore
\begin{eqnarray}
N(S)&\geq& [\rho^2- q^2-2(\rho + 3 \hat{p})q]\nonumber
\\&+&\sum_{i=1}^{3}[\rho^2-q^2-2(\rho+3\hat{p})q- p_i^2]A_i^2.
\end{eqnarray}
Accordingly, as in previous cases, the necessary and sufficient conditions to have $N(S)
\geq 0$ are
\begin{eqnarray}
\rho^2 &\geq& q^2 + 2 (\rho + 3 \hat{p}) q, \label{one} \\
\rho^2 &\geq& p_i^2 + q^2 + 2 (\rho + 3 \hat{p}) q \label{two},
\end{eqnarray}
where $i=1,2,3$. Finally, it is easy to see that if the condition (\ref{two}) is satisfied, then (\ref{one}) will be also satisfied.

\subsection{Summary}

Any realistic gravitational system needs to satisfy the weak, strong, and dominant energy conditions. When the source of gravitational field presents dissipative phenomena such as viscosity and heat flux, the weak energy condition and the strong energy condition are satisfied if
\begin{eqnarray}
\rho &\geq& 2 q ,\label{e1} \\
\rho + p_i &\geq& 2 q ,\qquad i = 1,2,3, \label{e2} \\
\rho + 3 \hat{p} &\geq& 4 q \label{e3}.
\end{eqnarray}
These inequalities force the energy density to be positive, and the gravitational field to be attractive. Finally, the dominant energy condition, which establish that the energy flux must be a future oriented timelike or null vector, is satisfied if 
\begin{equation}
\rho^2 \geq p_i^2 + q^2 + 2 (\rho + 3 \hat{p}) q, \qquad i = 1,2,3. \label{e5}
\end{equation}
When the heat flux vanishes, the above conditions reduce to those presented by E. Poisson in his classic book {\em "A relativist's toolkit : the mathematics of black-hole mechanics"} \cite{2004rtmb.book.....P}. It is worth mentioning that the energy conditions we present are obtained without considering the Eckart approximation and so they are valid in general, that is, the calculations are framework independent.

\section{Conclusion}
\label{sec:conclusions}

In this paper, we have written the energy-momentum tensor for a viscous fluid with heat flux in a comoving tetrad in which the stress tensor is diagonal and the heat flux vector has arbitrary components. Under the Eckart approximation of small velocity gradients, we have related the isotropic pressure and the principal stresses with the viscosity coefficients, the shear tensor, and the expansion factor. This relations allow us to understand, in a simpler way, the relation between the dissipation phenomena and the anisotropy (or isotropy) properties of the stress tensor. We have particularly shown that the principal stresses in the fluid are equal to the local equilibrium pressure when the viscosity coefficients vanished, or when the fluid is static (CASES 1 and 2). In both cases the viscosity effects are zero and the energy is not dissipated through internal friction processes. Nevertheless, as it was established in CASE 3, when $\eta=0$ and $\zeta\neq 0$, the stress of the fluid becomes isotropic but the bulk viscosity does not cancel, so the term $\zeta \Theta h^{\alpha\beta}$ in the energy-momentum tensor, due to the bulk movement of the fluid, contributes in the energy dissipation. Otherwise, if the bulk viscosity vanishes but the shear one does not, as in CASE 4, the stress of the fluid will become anisotropic and the energy will be dissipated through internal friction.

On the other hand, using the shear tensor definition (\ref{shear}) we have shown the spatial projections of $\sigma^{\alpha\beta}$ in terms of the extrinsic curvature of the normal surface to $u^{\alpha}$, and in terms of the Lie derivative of the metric tensor. These alternative ways of writing (\ref{relat1}) and (\ref{stresses}) allowed us to relate the principal stresses with the geometric properties of the spacetime. Moreover, when the diagonal elements of the extrinsic curvature are equal to $\Theta/3$ (CASE 5), the stress in the fluid is isotropic and the energy is dissipated through bulk viscosity. Finally, using the symmetries of the spacetime, it is found that when the fluid velocity vector is a linear combination of the Killing vectors, the principal stresses are equal and the isotropic pressure reduces to the local equilibrium pressure. If in addition, $q^{\alpha}=0$ then the fluid will be perfect. Hence, a connection between the dissipation phenomena and the symmetries of the spacetime is established.

Now, taking into account that any realistic source of gravitational field needs to satisfy some energy conditions, we have computed them for a viscous fluid with heat flux, whose energy-momentum tensor is written in the form (\ref{primertab}). We also consider the energy density and the energy flux density vector measured by an arbitrary observed, and then apply the method used by E. Poisson \cite{2004rtmb.book.....P} in order to find the energy conditions in terms of the thermodynamical properties of the fluid ($\rho$, $p_{1}$, $p_{2}$, $p_{3}$, $q$). The weak energy condition, that makes the energy density to be positive, and the strong energy condition, that ensures the attractive nature of the gravitational field, are both satisfied if the inequalities (\ref{e1}), (\ref{e2}), and (\ref{e3}) are simultaneously satisfied. Finally, the energy flux density vector is timelike and future oriented (dominant energy condition) if (\ref{e5}) is satisfied. These conditions were found following a simple algebraic process in which we have not needed to introduce any approximation or restriction on the energy-momentum tensor; hence, the conditions are general and they can be applied to any source of gravitational field in which the electromagnetic fields are not presented.

\begin{acknowledgements}
O. M. P. wants to thanks the financial support from COLCIENCIAS and Universidad Industrial de Santander. F.D.L-C gratefully acknowledges the financial support from COLCIENCIAS (Programa Es Tiempo de Volver) and Vicerrector\'ia de Investigaci\'on y Extensi\'on,  Universidad Industrial de Santander, grant number 1822. G. A. G. was supported in part by VIE-UIS, under Grants No. 1347 and No. 1838, and by COLCIENCIAS, Colombia, under Grant No. 8840.
\end{acknowledgements}



\begin{thebibliography}{10}

\bibitem{2013rehy.book.....R}
L.~{Rezzolla} and O.~{Zanotti}.
\newblock {\em {Relativistic Hydrodynamics}}.
\newblock Oxford University Press, Oxford (2013).

\bibitem{2010nure.book.....B}
T.~W. {Baumgarte} and S.~L. {Shapiro}.
\newblock {\em {Numerical Relativity: Solving Einstein's Equations on the
  Computer}}.
\newblock Cambridge University Press, Cambridge, UK (2010).

\bibitem{2003CQGra..20R.105A}
N.~{Andersson}.
\newblock {Topical review: Gravitational waves from instabilities in
  relativistic stars}.
\newblock {\em Class. Quant.
Gravity}, 20, R105–R144 (2003).

\bibitem{2007LRR....10....1A}
N.~{Andersson} and G.~L. {Comer}.
\newblock {Relativistic Fluid Dynamics: Physics for Many Different Scales}.
\newblock {\em Living
Rev. Relativ.}, 10(1) (2007)

\bibitem{2015PhRvL.115m2301R}
S.~{Ryu}, J.-F. {Paquet}, C.~{Shen}, G.~S. {Denicol}, B.~{Schenke}, S.~{Jeon},
  and C.~{Gale}.
\newblock {Importance of the Bulk Viscosity of QCD in Ultrarelativistic
  Heavy-Ion Collisions}.
\newblock {\em Phys. Rev. Lett.}, 115(13), 132301 (2015).

\bibitem{1984ucp..book.....W}
R.~M. {Wald}.
\newblock {\em {General relativity}}.
\newblock University of Chicago Press, Chicago (1984).

\bibitem{0264-9381-5-10-011}
C~A Kolassis, N~O Santos, and D~Tsoubelis.
\newblock Energy conditions for an imperfect fluid.
\newblock {\em Class. Quant.
Gravity}, 5(10), 1329 (1988).

\bibitem{2004sgig.book.....C}
S.~M. {Carroll}.
\newblock {\em {Spacetime and geometry. An introduction to general
  relativity}}.
\newblock Addison Wesley, San Francisco, CA (2004).

\bibitem{1973lsss.book.....H}
S.~W. {Hawking} and G.~F.~R. {Ellis}.
\newblock {\em {The large-scale structure of space-time.}}
\newblock Cambridge University Press, Cambridge, UK (1973).

\bibitem{1973grav.book.....M}
C.~W. {Misner}, K.~S. {Thorne}, and J.~A. {Wheeler}.
\newblock {\em {Gravitation}}.
\newblock W. H. Freeman and Company, San Francisco,
CA (1973).

\bibitem{Maartens:1996vi}
Roy Maartens.
\newblock {Causal thermodynamics in relativity(1996)}.
\newblock arXiv:astro-ph/9609119.

\bibitem{2004rtmb.book.....P}
E.~{Poisson}.
\newblock {\em {A relativist's toolkit : the mathematics of black-hole
  mechanics}}.
\newblock Cambridge University
Press, Cambridge, UK (2004).

\bibitem{1985fcgr.book.....S}
B.~F. {Schutz}.
\newblock {\em {A First Course in General Relativity}}.
\newblock Cambridge University Press, Cambridge, UK
(2000).

\bibitem{2000ifd..book.....B}
G.~K. {Batchelor}.
\newblock {\em {An Introduction to Fluid Dynamics}}.
\newblock Cambridge University Press, Cambridge, UK
(2000).

\bibitem{1959flme.book.....L}
L.~D. {Landau} and E.~M. {Lifshitz}.
\newblock {\em {Fluid mechanics}}.
\newblock Pergamon Press, Oxford (1959).

\bibitem{1940PhRv...58..919E}
C.~{Eckart}.
\newblock {The Thermodynamics of Irreversible Processes. III. Relativistic
  Theory of the Simple Fluid}.
\newblock {\em Phys. Rev.},  58, 919–924 (1940).

\bibitem{2010CQGra..27t5012C}
P.~{Cerd{\'a}-Dur{\'a}n}.
\newblock {Numerical viscosity in hydrodynamics simulations in general
  relativity}.
\newblock {\em Class. Quant.
Gravity }, 27(20), 205012 (2010).

\bibitem{1985JMP....26.2881M}
D.~P. {Mason} and M.~{Tsamparlis}.
\newblock {Spacelike conformal Killing vectors and spacelike congruences}.
\newblock {\em J. Math.
Phys.}, 26, 2881–2901 (1985).

\end{thebibliography}


%
%

\end{document}